\documentstyle[mncite,psfig]{mn}
\journal{Sussex preprint SUSSEX-AST 96/7-3, astro-ph/9607083}

\def\et{\em et al. \em}
\def\msun{M_{\odot}}
\def\gte{\,\lower.6ex\hbox{$\buildrel >\over \sim$} \, }
\def\lte{\,\lower.6ex\hbox{$\buildrel <\over \sim$} \, }

\title{Residual gas expulsion from young globular clusters}

\author[S.P.Goodwin]{Simon P. Goodwin\\
            Astronomy Centre, University of Sussex, 
            Falmer, Brighton BN1 9QH, UK.}

\date{}

\begin{document}

\maketitle

\begin{abstract}

The results of $N$-body simulations of the effects of the expulsion of 
residual gas (that gas not used in star formation) from very young 
globular clusters is presented.  Globular clusters of a variety of initial 
masses, Galactocentric radii, concentration and initial mass function 
slope with star formation efficiencies of $\lte 50\%$ were simulated.  The 
residual gas was expelled by the action of massive stars in one of 
three idealised ways: gradually by their UV flux and stellar winds; gradually 
by the input of energy by supernovae; and in a 'supershell' expanding from the 
cluster centre.  The clusters were compared shortly after the gas expulsion 
with the results of Chernoff \& Shapiro (1987) to estimate whether they 
would survive for a Hubble time.  It is found that the expulsion of 
$\gte 50\%$ of a globular clusters mass in a short period of time 
considerably affects the structure of the cluster.  However, many 
clusters are estimated to be able to survive with reasonable initial 
conditions, even if their star formation efficiencies are possibly as low 
as 20\%.  It is found that the central density 
required within a proto-globular cluster at star formation in order for it 
to survive at a given Galactocentric radius is independent of the mass of 
stars in the cluster.  For globular clusters in the inner few kpc of the 
Galaxy this value is found to be around $10^3 \msun$ pc$^{-3}$, falling as 
Galactocentric radius increases.  This value is similar to the central 
densities found in giant molecular clouds in the Galaxy today.   It is 
suggested that a globular cluster could reasonably form with that  
central density with a star formation efficiency of $\approx 40\%$ and 
and initial mass function slope $\alpha \approx 3$.  

\end{abstract}


\section{Introduction}
\label{sec:intro}

This paper reports the results of an $N$-body simulation of the early 
stages of the evolution of globular clusters.  These simulations differ 
from previous models in that they include the residual gas (that 
gas not used in star formation) left in the cluster immediately after 
their formation. The simulations model the dynamical effects upon the cluster 
of the expulsion of this gas and the implications for the cluster's 
survival.  The aim of these simulations is to constrain in a parameter 
space containing star formation efficiency, initial mass function, 
Galactocentric radius and 
the initial spatial and energy distributions of the stars the 
conditions that will result in a long-lived globular cluster.
  
The observed properties of globular clusters and candidate globular 
clusters provide a number of strong  
constraints upon the initial conditions in those clusters,  
which may then be applied to theoretical models.  Possibly the most 
important observational result from which it is possible to constrain these 
initial conditions is that relating to the chemical composition of 
globular clusters. Galactic globular clusters, with the important exceptions 
of $\omega$ Cen and M22, are extremely chemically homogeneous with 
differences in their internal metallicities of only about one tenth of a dex 
 (Fahlman, Richer and VandenBerg 1985 and Kraft \et 1992).  
This implies that the stars in globular clusters 
formed from a well-mixed gas cloud in the proto-Galaxy 
in a single burst of star formation.  This star formation 
may have been internally induced 
(Fall and Rees 1985, Harris \& Pudritz 1994 and Brown, Burkert \& Truran 1991,
 1995) or caused by external 
effects such as cloud-cloud collisions (Lin \& Murray 1991, Murray \& Lin 
1992, Kumai, Basu \& Fujimoto 1993 and Lee, Schramm \& Mathews 1995) or by 
the interaction of shock waves  
with clouds (Shapiro, Clocchiatti \& Kang 1992).  Notwithstanding the 
inducing mechanism, the star formation must have been 
rapid and efficient.  Star formation must be completed before the massive 
stars could ionise the residual gas and impede further star formation, as the 
ionisation of the gas will prevent the condensation of further stars 
from the gas in the proto-cluster.  Further 
to this condition, the  remaining gas in the cluster  must have 
 been expelled from the cluster.  This expulsion must have occurred 
in order to prevent another generation of stars forming which would have had 
a higher metallicity due to the 
enrichment of the residual gas by the supernovae of the massive stars 
within the cloud.  Hence the $10^{6}$ to $10^{7}$ yr lifetime 
of the massive stars gives an upper limit to the star formation timescale
 of the entire cluster (Lin \& Murray 1991).  The loss of the gaseous component of the cluster will reduce the binding energy of the cluster making 
it more susceptible to disruption.  The expulsion of the residual gas 
is presumed to be driven by the rapid evolution of the most massive 
stars.  The expulsion is assumed to be caused either by the energy released 
into the residual gas by supernovae or the cumulative effect of ionisation 
and stellar winds from these stars. 

Presuming that a globular cluster survives the loss of its  
residual gas, two further processes act upon the cluster which may  
disrupt it.  Firstly, the cumulative mass loss caused by the stellar evolution 
of individual stars within the cluster will lower the mass of 
the cluster over a long timescale.  
In addition, the evaporation of stars from the cluster will  
reduce the total cluster mass. Evaporation may be  caused by  the 
action of the galactic tidal field  or via relaxation where stars involved 
in energetic two-body encounters will be pushed into  
the high-velocity tail of the Maxwellian velocity  distribution where 
they reach escape velocity and leave the cluster.  
There have been many previous simulations of the evolution of globular 
clusters, although none of these simulations have included any residual gas 
loss.  The early work using $N$-body codes, the Fokker-Planck equation 
and hydrodynamic codes is summarised in Spitzer (1987) and Elson, Hut \& 
Inagaki (1987).
 More recent work has been primarily using the Fokker-Planck 
equation due to the large number of particles that can be included in those
 simulations (Weinberg
\& Chernoff 1988 and Chernoff \& Weinberg 1990).  Additionally, 
Chernoff \& Shapiro (1987) used a King model based simplified evolutionary 
calculation following the evolution of basic King model parameters.  It 
is to these calculations that the $N$-body results in this paper will be 
compared. 

The reasoning for using an $N$-body code in preference to a Fokker-Planck 
simulation is two-fold.  Firstly, the $N$-body method makes no a priori 
assumptions about the dynamical evolution of the system.  Secondly, 
an $N$-body approach allows the effects of an variable external potential 
to be very easily modelled.  The 
 code is based upon Aarseth's nbody2 code (Aarseth 1985) with  
an initial mass function for particles.  The code included  
stellar evolution and a variable external potential to simulate the 
residual gas loss.  $N$-body simulations of open clusters 
including some gas loss have 
concentrated upon open star clusters (Lada, Margulis \& Dearborn 1984, 
Terlevich 1987) due to the particle number limitations of $N$-body 
codes in comparison to the Fokker-Planck approximation.  Recently 
$N$-body codes have been applied to the evolution of globular clusters in 
a statistical approach 
as in Fukushige \& Heggie (1995) where the results of $N$-body and 
Fokker-Planck simulations were found to be qualitatively 
similar (this approach been analysed by Giersz \& Heggie 1994a, 1994b). 
The $N$-body approach has been useful in discovering a number of properties of 
the evolution of large $N$ systems such as core collapse 
(Aarseth \& Lecar 1975).
It has also been used to test weak-scattering Fokker-Planck 
models (Aarseth, H\'{e}non \& Wielen 1974) and the tidal evolution of 
globular clusters (Oh, Lin \& Aarseth 1992). 

The simulations were run to cover $50<T<100$ Myr, that is the timescale 
for the loss of the residual gas (approximately the life time of the 
most massive stars) and a 'settling down' period afterwards.  
Following the evolution for such a relatively short period of time 
reduces the computing time and the problems of an $N$-body calculation.

In order to estimate the final fate of the cluster (whether it will 
survive or disrupt) the condition of the cluster after the 
simulation is compared to the results of Chernoff \& Shapiro (1987), 
hereafter CS.  The simulations of CS, based upon King models, 
give the final fate of a cluster for a wide range of initial  
Galactocentric radii, initial mass functions and concentrations.  The range 
of these initial conditions is far wider than those of any other 
simulation.  They are 
found to have broad agreement with Fokker-Planck (eg. Chernoff \& Weinberg 
1990) and $N$-body (Fukushige \& Heggie 1995) simulations of globular 
cluster evolution, although it should be noted that Fukushige \& Heggie 
(1995) do conclude that the types of Fokker-Planck simulations used to 
model globular clusters may be 'quantitatively very unreliable'.

\section{Initial Conditions}

The initial conditions are set to be fairly similar to clusters that CS 
estimate will survive a Hubble time.

\subsection{Tidal Cutoff Radius}

The outer edge of the cluster is taken to be the King (1962) 
tidal radius $r_{\rm t}$ (the point at which the density is assumed to vanish),
 given by

\begin{equation}
r_{\rm t} = R_{\rm G} \left( \frac{M_{\rm cl}}{M_{\rm G}(R_{\rm G})} \right)^{^{1}\!/_{3}}
\end{equation}

\noindent where $M_{\rm cl}$ is the mass of the cluster, $R_{\rm G}$ is the 
 Galactocentric distance of the cluster and $M_{\rm G}(R_{\rm G})$ is the mass of the Galaxy interior to $R_{\rm G}$.  $M_{\rm G}(R_{\rm G})$ is derived 
from the rotation curve of the Galaxy using $M_{\rm G}(R_{\rm G}) = R_{\rm G}v_{\rm c}^2 / G$ where $v_{\rm c}$ is the circular rotation velocity 
($\approx 220$ km s$^{-1}$) and 
$G$ is the gravitational constant.  This approximation is used for $R_{\rm G} > 2$ kpc.  Such a model is used instead of more complex Galactic potentials 
as globular clusters are very old objects which formed during the early 
stages of Galactic evolution when the Galactic potential was probably 
very different from that today.  If the Galaxy was still in the process 
of accreting mass when globular clusters were formed this would produce 
a correspondingly lower tidal radius at the time of formation.  However, 
as $r_{\rm t} \propto M_{\rm G}(R_{\rm G})^{-1/3}$ this effect may not 
be too important.

This form of the tidal radius is an idealised 
case which is spherically symmetric.  In real clusters the shape of the equipotential surfaces will be elongated in the radial direction 
pointing towards the Galactic centre (see Spitzer 1987 and Heggie \& 
Ramamani 1995).  For globular clusters 
with large Galactocentric distances ($\geq 6$kpc) the extent of this 
asymmetry is small and the distance of the tidal boundary from the 
cluster centre is large in comparison to the half-mass radius. The 
simplification to a spherically symmetric boundary will not greatly 
effect the rate of stellar evaporation in the short 
periods of time covered by these simulations where the effects of a more 
complex treatment of tidal forces, such as those used in Terlevich(1987) and 
Fukushige \& Heggie (1995), should not be important.  

For present day globular clusters this tidal radius is normally in the range 
$50 < r_{\rm t}< 100$ pc (using Galactocentric radii and masses from Chernoff 
\& Djorgovski 1989).  If a star passes beyond
 this distance (ie. its energy is above the escape energy of 
$-GM_{\rm cl}^2/r_{\rm t}$) it is assumed to be outside of the gravitational 
influence of the cluster and  to leave the cluster forever.  This is 
not necesserally the case, but is usually accepted 
as a first approximation in models of globular clusters (Spitzer 1987). 
 This mass loss due to evaporation of stars from 
the cluster will, of course, reduce the tidal radius allowing further 
evaporation to occur more easily.  

\subsection{Initial Mass Function and Stellar Evolution}

The initial mass function (IMF) is taken to be a power law of the form 
 
\begin{equation}
N(M) \propto M^{-\alpha}
\end{equation}

\noindent where this corresponds to the Salpeter (1955) IMF for the solar 
neighbourhood when $\alpha=2.35$. Such forms for the IMF are used in 
virtually all simulations of globular clusters (for example CS, Chernoff 
\& Weinberg 1990 and Fukushige \& Heggie 1995).  Three values of $\alpha$ 
are used in the models of $2.35$, $3.5$ and $4.5$ which cover the most 
acceptable probable initial values of the IMF of a globular cluster and to 
enable simple comparisons of these simulations to CS.  Chernoff \& Weinberg 
(1990) have shown that no cluster with an IMF slope of $\alpha =1.5$, or 
lower, is able to survive the large amount of mass loss due to stellar 
evolution in the first $5$Gyr of its life so such low IMF slopes were not 
considered.  
The IMF is implemented in the code by assigning each particle a different mass 
according to the slope of the IMF.  Each of these particles 
represents a collection of similar mass stars close together in phase space.  
Each particle contains from a few tens to a few thousands of stars, and 
the average mass of particles in the simulations can 
range from tens of solar masses to tens of thousands of solar masses 
depending upon the initial total mass of the cluster being simulated and the 
number of particles used to represent the cluster.  The range of initial 
masses studied in this paper ranges from $10^{4}\msun$ to $10^{6}\msun$, 
normally using $1000$ particles.  This is obviously a gross approximation 
but allows the code to simulate the overall evolution of the cluster in a 
simple way.

The actual upper mass limit of the stars in the cluster is taken to be 
$12\msun$.  As stated above each particle in the simulation represents 
far more than one star, as stars of above $12\msun$ are very rare, it would 
be inappropriate to assign to them more than one individual particle.

The lower mass limit of the IMF is taken to be $0.15 \msun$ corresponding 
to the observed downturn of the mass function in globular clusters 
(Paresce, De Marchi \& Romaniello 1995).    

One consequence of this treatment of the IMF is that the particle 
representing the highest mass stars is far larger than other particles 
in the simulation.  In order to see if this could effect the 
results a number of simulations were run where this one large particle was 
split into 10, or more, different particles which evolved, in total, like 
the larger particle.  The results of simulations 
with one large particle and several smaller particles were virtually 
identical.  The use of such a large particle is justified as it does not 
survive for long enough before evolving to have a significant effect 
upon the dynamics of the other particles through two-body encounters.

Mass loss due to stellar evolution plays a very important part in 
the evolution of a globular cluster, especially in the early stages of 
its life when the rapid evolution of massive stars will substantially 
alter the mass of the cluster and correspondingly lower the tidal 
radius (Chernoff \& Weinberg 1990).  
The effects of stellar mass loss have been included by the fitting of a 
simple straight line to the end times of stellar evolution calculated by 
Maeder \& Meynet (1988) for various masses of Solar  
metallicity stars

\begin{equation}
{\rm Log_{10}} \left( \frac{M}{M_{\odot}} \right) = 1.524 - 0.370{\rm Log_{10}} 
\left( \frac{T}{\rm Myr} \right)
\end{equation}

The majority of globular clusters are of lower than solar metallicity and 
these stellar evolutionary times are calculated for solar metallicity.  
However, as the only evolutionary times of interest are those of high mass 
stars these evolutionary times are expected to be close  to those for low 
metallicity stars.   
  
Once a star has reached its final age it evolves to the appropriate end 
state for a star of that mass. 
Stars are divided into three mass categories each of which leave a 
different stellar remnant at the end of their evolutionary times.  Stars of 
$M_{*} > 8\msun$ become type II supernovae, leaving behind a $1.4\msun$ 
neutron star.  For stars of intermediate mass, $4 < M_{*}/\msun < 8$ 
 two end states are possible.  The first is that the star becomes a type 
I$^{1}\!/_{2}$ supernova which catastrophically disrupts leaving no 
remnant (Iben and Renzini 1983).  Alternatively 
the star may evolve into a white dwarf of mass $1\msun$.  It is not 
known how important each of these processes are relative to each other, so 
both end states are used in the code. The loss of mass from the cluster 
is obviously greater in the first case but due to the steep 
slopes of the IMF used (hence the relatively small number of 
intermediate mass stars) this difference is quite low.  Low mass stars of 
$1 < M_{*}/\msun < 4$ all evolve into white dwarfs of mass 
 $0.58 + 0.22(m-1)$, where $m$ is the initial mass of the 
star, all in solar masses (Iben \& Renzini 1983).  Because of the short 
timescale followed by the code (the evolutionary timescale is only followed 
to a maximum of 100 Myr) only high mass stars and maybe a very few 
intermediate mass stars have the time to evolve.

The evolution of each particle is not instantaneous. Each particle is 
assumed to contain stars of a similar mass, but not all exactly the same 
mass. The range of masses contained in each particle fills half of the 
gaps between it and the masses of particles on either side in the mass 
spectrum.  This removes the wide spacing in masses at the upper end of 
the mass spectrum, especially when the IMF slope is high.  Thus the total 
change in mass of the particle from $N_{*}$ stars to $N_{*}$ lower mass 
remnants will not occur simultaneously. This is due to the dependence of 
lifetime upon stellar mass.  This difference is small for high mass stars, 
but can become very pronounced for stars of a low mass (see equation (3)).  
The code spreads the change in mass of a particle over a few Myr, this has 
the advantage of not allowing drastic instantaneous changes which are 
obviously not realistic (especially when a particle whose average stellar mass 
is $12\msun$ becomes a particle of neutron stars of mass $1.4\msun$).

This multi-mass model of the IMF tends to clump similar mass stars in 
phase space.  To see if there were significant differences between this 
method and one in which particles have a uniform mass and contain the 
full range of masses comparison simulations were made.  The two methods 
are qualitatively similar over the timescales of these simulations 
($\approx 50$ Myr).  Discrepancies occur in the numbers of particles that 
escape beyond the tidal radius.  The total mass of escaping particles is 
similar, but more particles are lost in the multi-mass case, the majority 
from the low-mass end of the IMF.  Single-mass simulation clusters 
also appear to have a slightly higher survivability.  The difference is small 
and probably unimportant in view of the qualitative nature of the 
results presented.

The mass lost from stars due to stellar evolution is treated in two 
ways depending upon whether the residual gas is still present in the 
cluster. Whilst the residual gas is still in the cluster, the mass 
 lost by stars is neglegable in 
comparison to the total mass of gas in the cluster and so can be ignored (even 
though the energy provided by this mass loss 
is the driving force behind much of the gas   
expulsion in the cluster).  After residual gas has left the cluster 
further mass lost through stellar evolution is assumed to leave the cluster immediately.  Even if the gas does not leave the cluster instantaneously, 
the timescales followed in the simulation would not allow significant 
amounts of gas to collect. 

\subsection{Initial Distribution of Stars and Gas}

Immediately after star formation a globular cluster will be composed of 
roughly equal proportions (to within an order of magnitude) of stars 
and gas.  The initial ratio of stellar mass to total mass determines the star 
formation efficiency (SFE) of the proto-cluster cloud. The initial distributions of both the stars and gas in the cloud 
are represented by a Plummer (1911) potential given by

\begin{equation}
\phi (r) = - \frac{GM}{(r^{2} + R_{\rm S}^{2})}
\end{equation}

\noindent where $M$ is the total mass of the residual gas or stars, 
$R_{\rm S}$ is the scale length of the potential and $G$ the gravitational 
constant. The SFE is then given by 

\begin{equation}
\eta = \frac {M_{\rm stars}}{(M_{\rm stars}+M_{\rm gas})}
\end{equation}

The initial positions and velocities have been determined using the 
technique described in detail in Aarseth, H\'{e}non \& Wielen 
(1974).  The stars are first distributed randomly in phase space in 
such a manner that they produce a Plummer distribution.  The postions 
and velocities can be scaled to produce a distribution of the 
desired scale length and virial ratio (see below).  The 
gas is represented by an external potential acting upon the star particles.  
The scale lengths for the stars and gas are equal for any model.  This 
represents an assumed dependence of star formation upon gas density.    
A Plummer potential was chosen for the initial conditions in preference 
to a King model due to the simple analytic form of the potential.  
The gas is assumed to always be a spherically symmetric distribution and the stars are assumed to have no effect upon the gas, ie. the viscosity 
is neglegable and the gas is expelled from the system before any 
gravitational effect from the stars can make itself manifest upon the 
distribution of the gas.  Any 
change in the gas potential is idealised (see section 2.4) so that it 
remains in a Plummer model to simplify the potential and force 
calculations.  Selecting the value of $R_{\rm S}$ correctly the Plummer 
model can be made very similar to a King model.

Particles are initially placed in the inner region of the cluster in a sphere 
of $\approx 10$ to 30 pc radius, depending upon the initial concentration 
of the cluster and its Galactocentric radius.  This positioning is also 
independent of the mass of the particle.  Stars are not assumed to be 
formed over an extended region of space, certainly not up to the tidal radius.

The initial velocity distribution of the stars is scaled according to the 
initial virial ratio, $Q$, of potential energy, $\Omega$, to kinetic 
energy, $T$ where $Q=T/\mid \Omega \mid =0.5$ corresponds to a system in 
virial equilibrium.  The virial ratio of the system is given by the 
equation summing over the number of particles in the simulation 

\begin{equation}
Q = \frac{ \sum_{i} m_{i}v_{i}^{2}}{  \sum_{i} m_{i} \sum_{j \neq i} 
\frac{Gm_{j} r_{ij}}{(r_{ij}^{2}+\epsilon^{2})^{^{3}\!/_{2}}} + \sum_{i} \frac{Gm_{i}M_{\rm gas}r_{i}}{(r_{i}^{2}+R_{\rm S(gas)}^{2})^{^{3}\!/_{2}}}}
\end{equation}

\noindent where $m_{i}$ is the mass of particle $i$, $M_{\rm gas}$ is 
the total mass of gas in the system, the radius of particle $i$ is $r_{i}$,  
the inter-particle distance between two particles $i$ and $j$ 
is given by $r_{ij}=|r_{i} - r_{j}|$, $\epsilon$ is the softening parameter 
(see below) and $R_{\rm S(gas)}$ is the scale length of the gas potential.

It is important to note that even clusters initially in virial 
equilibrium will not be in complete dynamical equilibrium.  It seems 
highly unlikely that after the star formation episode that the stellar 
component of the cluster 
would be in dynamical equilibrium, so some settling of stars into an 
equilibrium distribution after star formation would be expected, even if there 
were no gas or no gas expulsion.

After the mass loss episode in the cluster, the evolution will proceed 
along normal lines for a globular cluster with evaporation and stellar 
evolution competing against core collapse to decide the final fate of the 
cluster.

\subsection{Residual Gas and Mechanisms for gas loss}

As argued in the introduction, a globular cluster must expel all of its 
residual gas before another generation of stars is able to form in order 
to retain its extreme chemical homogeneity.  This mass loss will have 
an important effect upon the dynamics of the cluster possibly 
leading to its destruction.  The simplest application 
of the virial theorem to star clusters implies that for a SFE of less than 50\% 
a cluster initially in virial equilibrium cannot lose all of its 
residual gas and still remain bound.  More sophisticated simulations of 
open cluster dynamics show that a system may retain a bound core of stars 
with a SFE as low as 30\%  (Lada \et 1984). 

This expulsion of gas is assumed to be driven by the high mass 
($M_{*} > 8\msun$) stars in the cluster through their  
photoionisation of the surrounding medium by intense UV radiation 
(Tenorio-Tagle \et 1986), strong stellar winds and their final supernovae
 explosions (Dopita \& Smith 1986 and Morgan \& Lake 1989). These three 
different mechanisms are simulated to represent the alternative routes by 
which the gas may be lost. 

The first mechanism of gas expulsion modelled is that due to photoionisation 
and strong stellar winds (refered to throughout as gradual early gas 
expulsion).  Tenorio-Tagle \et (1986) used a hydrodynamic code to model 
the expulsion of gas from a cluster.  They showed that 
for relatively low masses of gas ($10^{3}$ to $10^{4} \msun$), the 
photoionisation caused by 100 O5 stars, each producing some 
$4$x$10^{52}$ ergs of UV radiation in their lifetime (Chiosi \& Maeder 1986), 
 may expel all of the gas within a cluster, while for higher 
gas masses only a small, inner region is ionised and no gas is lost. 
The timescale for this gas loss is only a few Myr starting around 4Myr 
after the formation of the massive stars. These gas masses are  
far lower than the masses of residual gas left in young globular 
clusters. However, a proto globular cluster will normally contain more 
than enough stars to expel the residual gas, at least by the action 
of supernovae (for a caveat see section 3.5).  The number of massive stars 
of above some mass $M_{\rm SN}$ in a cluster with an IMF of slope $\alpha$ 
is given by

\begin{equation}
N_{\rm SN} = M_{\rm cl} \frac{(\alpha - 2)}{(\alpha - 1)} \frac{(M_{\rm SN}^{-(\alpha - 1)} - 
M_{\rm up}^{-(\alpha - 1)})}{(M_{\rm low}^{-(\alpha - 2)} - M_{\rm up}^{-(\alpha - 2)})}
\label{eqn:nsn}
\end{equation}

\noindent where $M_{\rm low}$ and $M_{\rm up}$ are the upper and lower mass limits 
of the IMF respectively and $M_{\rm cl}$ is the total mass of the cluster.  
Using $M_{\rm SN}=8\msun$, $M_{\rm up}=15\msun$ (note that this is the highest 
mass of star assumed present in the cluster, rather than the 
mass of star represented by the largest particle) and 
$M_{\rm low}=0.15\msun$ then $\alpha = 3.5$ will give $N_{\rm SN} \approx 
150$ for a cluster where $M_{\rm cl}=10^{6}\msun$. An additional 
source of energy not included in Tenorio-Tagle \et is that 
each O5 star will also add around $10^{49}$ ergs into the 
interstellar medium through strong stellar winds at typical globular 
cluster metallicities (Kudritzki, Pauldrach \& Puls 1987). It 
may well, then, be reasonable to assume that in some cases mass loss can 
be driven solely by these two processes.  This mechanism for 
mass loss will begin only a few Myr, at most, after the end of star 
formation and expel the majority of the residual gas before the first 
supernovae explode. 

This type of mass loss is modelled in the code by reducing the mass of gas 
$M_{\rm gas}$ in the Plummer potential gradually with time.  The loss of the 
gas begins after $\approx 4$ Myr and continues at a constant mass loss rate 
(based upon those found in Tenorio-Tagle \et 1986) until no gas is left within 
the cluster (cf. Lada \et 1984).  During the gas expulsion episode the 
scale length of the gas remains constant (and equal to that of the stars).
 
The second and third mechanisms of gas loss are both via the supernovae 
explosions of massive stars.

The second mechanism is similar to the 
loss of gas via photoionisation and strong stellar winds above in that 
the mass of gas in the gas potential is gradually reduced.  However, this 
reduction begins at a later time as it is caused by the supernovae of 
massive stars expelling the gas (refered to throughout as gradual late 
residual gas expulsion).  It is partly based upon the 
calculations of the effects of supernovae on gas clouds (Dopita \& Smith 
1986 and Morgan \& Lake 1989) as to the number of supernovae required to 
disrupt a certain mass of gas. The main difference between this mechanism 
and the first is that the onset of mass loss is delayed until the most 
massive stars ($\approx 15\msun$) and below 
reach the end of their lives and go supernova ($\approx 10$ Myr or more).

Both gradual mechanisms are assumed to have timescales of gas expulsion 
that are independent of the mass or Galactocentric radius of the 
cluster.  There seems little reason to suspect that Galactocentric radius 
would effect the expulsion timescale other than that the gas would 
require longer to escape beyond the tidal radius of high $R_{\rm G}$ 
clusters.  This is assumed to be unimportant as most stars are initially 
well within the tidal radius.  There would probably be 
a dependence of expulsion rate with cluster mass.  However, the mechanics of 
gas expulsion are very poorly understood and so an independence is 
assumed.  This assumption is used as a first approximation as higher 
mass clusters will contain more high mass stars which expel the gas.

The third mechanism is a highly idealised simulation of the effects on a 
star cluster of mass ejection via a 'supershell' such as those observed 
in  OB associations in the Galactic disc.  These supershells are 
formed by the merging of 
many supernova shock fronts into one large and powerful shock (McCray \& 
Kafatos 1987, McCray \& Mac Low 1988).  A moderate number of supernovae in 
the central regions of a proto-cluster cloud will be able to form a 
supershell to force the gas out of the gravitational influence of the cluster 
(Brown \et 1995).  A supershell will have a radius $r_{\rm shell}$ 
dependent upon the rate of supernova events $\dot{N}$. $\dot{N}$ is 
calculated from the total number of supernovae events in the cluster, 
given by equation(~\ref{eqn:nsn}), assuming that the events are spread evenly 
throughout the whole time that supernovae are occuring. The radius is 
given by equation 9 in Brown \et (1995) as

\begin{equation}
r_{\rm shell} = \left( \frac{3}{10\pi} \frac{E\dot{N}t}{P_{\rm ext}} \right) 
^{^{1}\!/_{3}}
\end{equation}

\noindent  where $E$ is the energy of each supernova (taken 
to be $10^{51}$ ergs), $P_{\rm ext}$ is the external pressure of the gas in the 
cloud and $t$ is 
the time since the first supernova . The supershell is approximated 
in the code by assuming that the shell is spherically 
symmetric and that all matter interior to $r_{\rm shell}$ has been swept into 
the shell.  The finite size of the shell is also neglected. Stars with 
$r_{*} < r_{\rm shell}$ feel no external potential due to the gas while stars 
with $r_{*} > r_{shell}$ feel the force due to the gas as if there were no 
supershell (applying Newton's First Theorem).  Once $r_{\rm shell} > r_{\rm t}$ then 
all of the gas is assumed to be lost into the intra-cluster medium.  Again, 
the onset of this mechanism is delayed until the most massive stars 
go supernova.

\begin{figure*}
\centerline{\psfig{figure=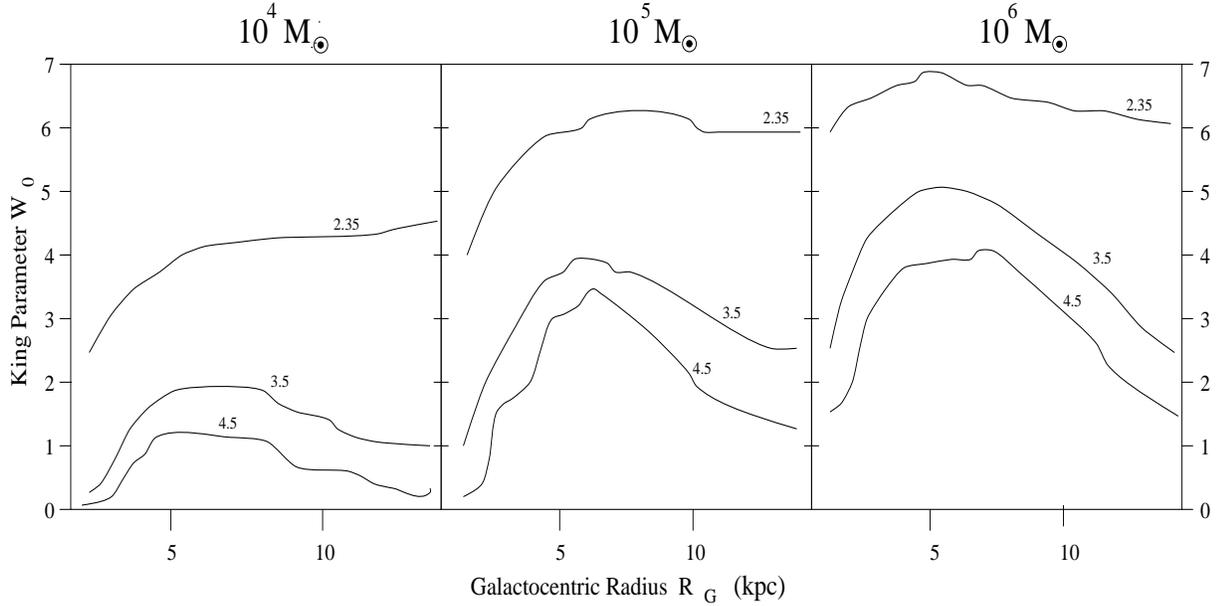,height=8.0cm,width=16.0cm,angle=270}}
\caption{A compressed version of fig. 4 from Chernoff and Shapiro 
(1987).  The three boxes show the border-line value of the central potential 
parameter $W_0$ of a cluster with the specified initial stellar mass at 
some Galactocentric radius.  The 
three lines indicate (from the uppermost in all three cases) the slope 
of the IMF $\alpha=2.35$, 3.5 and 4.5.  Clusters above the line will 
survive and those below will be disrupted.}
\label{figure:cs}
\end{figure*}

While $r_{\rm shell} < r_{\rm h}$ the assumption of a central source of 
energy from supernovae is not good.  The dynamics of the shell and the 
consequent effect upon the stellar dynamics of the cluster are not well 
modelled. This mechanism only has the effects of a very rapid depletion of 
the residual gas in the inner regions. However, this is only true for a 
short period (normally less than a crossing time) and so are not expected to 
be of great importance.  This model is more concerned with the effects of 
a supershell upon the dynamics of outlying particles which may wait a far 
longer before the shell passes their positions.  

These expulsion timescales are often in excess of those used by Lada \et 
(1984) whose maximum time for the expulsion of gas was set to be four 
crossing times.

These three expulsion mechanisms are treated in isolation in this paper.  In 
reality the situation will be far more complex and some aspect of all 
three mechanisms, and mechanisms not considered within this paper, will 
conspire to expel the residual gas.  The relative importance of the 
mechanisms, however, is unknown and it may be that any one (or none) is by 
far the most dominant. 

\subsection{Computational Aspects}

Standard $N$-body units were used from Heggie \& Mathieu (1986) to scale 
such that $M_0=G=1$ and 
$E_0=-1/4$, where $M_0$ is the initial mass of the particles (ie. the initial 
stellar mass of the cluster) and $E_0$ is the initial energy of the 
particles.  Conversion to units of time (in Myrs) for the treatment of 
stellar evolution was made using the relationship

\begin{equation}
T_{\rm c} = M^{5/2}_0 / (2 \, \vert E_0 \vert)^{3/2} = 2\sqrt 2 U_t  
\end{equation}

\noindent where $U_t$ is the unit of time within the code.

The softening, $\epsilon$, in the code was set to be of order the 
inter-particle 
distance in order to reduce the effects of two-body relaxation.  For the 
duration (in 'real' time $\approx 50$ Myr) of the simulations the effects 
of two-body relaxation (see Spitzer 1987)  
would be neglegable.  This was a primary reason for the choice of such 
short durations as it minimises one of the major problems associated with 
an $N$-body simulation, especially with the relatively small number of 
particles (1000) used in these runs. 

The code was run on a Sun 10 Workstation at the Astronomy Centre, University 
of Sussex.  A typical run of 1000 particles over 100 crossing times 
took approximately 2 hours.

\subsection{Comparison with Chernoff \& Shapiro}

Given an initial mass, IMF, Galactocentric distance and 
the central potential parameter of the King model (for details of King 
models see King 1966) the results of CS enable 
the end state of the cluster to be estimated.  The end states possible in 
CS are disruption, steady state King model, collapsing or core collapsed.   
In the context of this paper these possible states will be divided into 
two:  disruption and survival.  The initial conditions that lead to these 
two possible states are illustrated in fig.~\ref{figure:cs}.

After the gas loss episode and a short settling period (simulations are 
normally run for 50 Myr) in the $N$-body simulations the continuing 
evolution of the cluster is assumed to be as CS.  The state of the 
$N$-body simulation after this period is then compared to the lowest King model 
that CS say will survive.  If the structure of the simulation is such that 
it is more concentrated and more bound that the lowest King model then 
it is assumed that the cluster would be able to survive.  A comparison is 
made of the mass distributions and velocity dispersions of the simulation 
and King model to asses the survivability.

The comparison of young globular clusters to King models appears justified 
as observations show that young LMC clusters have a luminosity profile 
remarkably close to that of a King model (Elson 1991, Elson, Fall \& 
Freeman 1987).

It should be noted that within the context of the CS paper the survival 
of clusters was maximised by the associated approximations.  CS also 
took no account of other destructive mechanisms, such as bulge and disc 
shocking, which will effect the evolution of globular clusters, Aguilar, Hut 
\& Ostriker (1988) discuss these, and other, mechanisms and their 
possible effects upon the globular cluster population of the Galaxy.  For 
these reasons, many more clusters than predicted in both CS and this 
paper may be disrupted over the course of a Hubble time.  

\section{Results}

In this section the results of the simulations are presented.  In section 
3.1 the general effects of residual gas loss upon a cluster and their 
dependence upon the environment of the cluster are investigated.  Section 3.2 
explores the three gas expulsion mechanisms themselves in more detail.  
Section 3.3 deals with the actual survivability according to CS of a wide 
variety of 50\% SFE clusters, while section 3.4 examines the effect of 
the initial virial ratio upon this survivability.  Finally, section 3.5  
looks at the survivability of clusters that have an SFE of less than 50\%.

Throughout this section particular attention is paid to clusters 
with an initial stellar mass of $10^5 \msun$.  This is due to the 
overwealming number of actual globular clusters that have approximately 
this mass.  

\subsection{The effects of residual gas expulsion}

The effect of residual gas expulsion on a globular cluster are initially 
independent of the chosen IMF of that cluster.  This result is as would 
be expected because of the relatively short timescale of the gas expulsion 
compared to other dynamical processes affecting the evolution of the 
cluster which would be dependent upon the mass spectrum of the stars (the 
amount of stellar mass lost is smaller with higher slopes to the IMF as the 
mass of high mass stars is obviously greater).  
In the few Myr covered by these simulations no dynamical 
processes such as relaxation, equipartion or mass loss due to the stellar 
evolution of anything other than very high mass stars should have any 
significant effect.  

The number of massive stars may well effect the gas loss 
rate, but in these simulations it has been assumed that there are enough stars 
to expel the gas and that this expulsion will occur on an arbiterally chosen 
timescale.  Only in the supershell mechanism is any account taken of the 
numbers of massive stars present in the cluster.  In this case the early 
stages of gas loss are relatively insensitive to changes in the number 
of massive stars present and so the effect upon the majority of stars 
(which populate the inner few pc) is effectively independent of the IMF.  In 
simulations run without stellar evolution at the three different IMFs the 
absence of the additional mass loss from the stellar evolution is not 
found to alter the bulk effects of the gas expulsion mechanism. 
However, the absence of stellar evolutionary mass loss will effect the 
survivability of the cluster as clusters with low values of $\alpha$ 
lose more mass and will have an extra disruptive influence upon them than 
clusters with higher $\alpha$.

\begin{figure}
\centerline{\psfig{figure=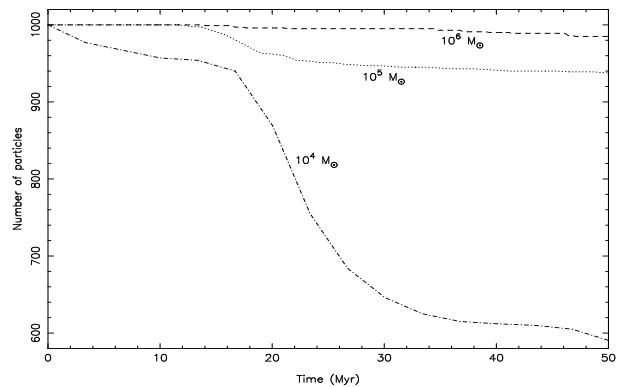,height=5.0cm,width=8.0cm,angle=270}}
\caption{Comparison of the change in the number of particles in the simulation  
(due to the escape of stars) with time in three otherwise 
identical clusters of $10^4 \msun$ (dot-dash line), $10^{5} \msun$ (dotted 
line) and $10^6 \msun$ 
(dashed line) illustrating the greater stability to residual gas loss of 
higher mass clusters.  All clusters initially had $R_{\rm S}$ = 3 pc and 
were in virial equilibrium at a Galactocentric distance of 5 kpc with 
gradual gas loss over 5 Myr starting 5 Myr after star formation.}
\label{figure:diffmass}
\end{figure}

The initial mass of the cluster is found to have a significant effect upon 
the effects of residual gas loss as shown 
in fig.~\ref{figure:diffmass}.  Note that the number of particles has been 
chosen instead of the cluster's stellar mass as it does not include the 
loss of mass due to stellar evolution, just the escape of particles 
beyond the tidal radius.  For otherwise identical clusters in 
IMF, Galactocentric radius, SFE, initial stellar and gas distribution 
and mass loss mechanism and timescale, higher mass clusters are found 
to be far more stable to gas loss than those of a lower mass.  Both the 
reduction in the binding energy of the stars and the expansion of the 
radial mass distribution are far less extreme in higher mass clusters.  
More massive clusters also lose fewer stars.  The reason for this 
dependence appears to lie in a very strong correlation between the number 
of crossing times over which gas loss occurs and the disruptive effect 
of that gas loss.  As the mass loss timescales for each mechanism 
are the same for all clusters, regardless of mass, then the 
gas loss occurs over more crossing times in higher mass clusters (see 
section 2.5).  Any dependence of expulsion timescale with cluster mass 
would be expected to increase the timescale in larger clusters (as $N_{\rm SN} 
\propto M$ but $\Omega \propto M^2$).  Longer expulsion timescales are 
less disruptive as the potential changes more slowly which would further 
increase the relative survivability of high mass clusters compared to 
lower mass clusters (as the number of crossing times over which gas loss 
would occur would be further increased).

The sudden increase in the rate of particle loss from the $10^4 \msun$ 
cluster at $\approx 18$ Myr is due to a large overflow of particles 
from the tidal radius.  This is caused as the tidal radius is smaller 
than the edge of the new equilibrium distribution of the cluster after 
mass loss.  The cluster expands to reach an equilibrium state and in doing 
so causes over 30\% of its particles to escape immediately.  The loss rate 
slows somewhat after 10 Myr, as all of the very energetic particles are 
lost, but still continues at a far higher rate than 
for the two more massive clusters until the destruction of the cluster.  This 
effect is self perpetuating in that the loss of particles (mass) causes 
the tidal radius to shrink even further and enhances the loss rate. 
The two larger 
clusters (with their correspondingly larger tidal radii) do not have this 
effect as fewer of their particles are able to reach this distance from the 
cluster core.  This effect results in the very rapid destruction of the 
$10^4 \msun$ cluster over $\approx 100$ Myr.

This effect is, to a certain extent, a result of the $N$-body simulation.  
In all masses of cluster 1000 particles have been used to simulate 
the cluster.  In the nbody2 code, the only place where the total actual 
mass of the particles (as opposed to their relative masses) is in the 
conversion from the code time units (in units of the crossing time) to 
physical time (in Myrs) see equation (9).  The system can be easily set-up in 
such a way 
that, in all but this conversion, the system is identical.  With this 
situation the higher stability of higher mass clusters is obvious after 50 Myr 
of physical time.  It appears that this effect may be 
representative of a real physical effect in actual clusters.

Further hydrodynamic simulations of gas expulsion  from 
such systems, similar to those of Tenorio-Tagle \et (1986), would be 
required to asess the true validity of the 
assumption that expulsion timescales in this cluster mass range are, indeed, 
as independent of initial cluster mass and tidal radius (and, even, the 
cluster IMF slope).

A dependence of the survivability of clusters 
upon Galactocentric radius is also found.  
Due to the dependence of tidal radius upon Galactocentric radius the 
amount of mass loss by overflow beyond the tidal radius stimulated by both 
the residual gas loss and the expansion of the cluster to a new 
equilibrium position is increased with decreasing Galactocentric radius, due 
to the linear dependency of $r_{\rm t}$ upon $R_{\rm G}$ (equation (1)).

\begin{figure}
\centerline{\psfig{figure=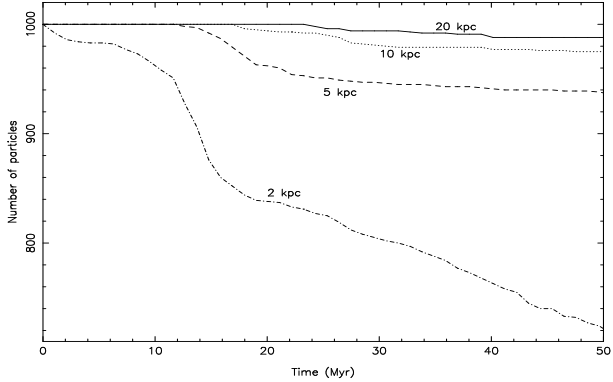,height=5.0cm,width=8.0cm,angle=270}}
\caption{Comparison of the change of the number of particles  
(from the escape of stars) with time in four otherwise 
identical clusters at Galactocentric distances of 2 kpc (dash-dot line), 
5 kpc (dashed line), 10 kpc (dotted line) and 20 kpc (solid line) showing 
the greater mass loss incurred at lower Galactocentric radii due to the 
smaller tidal radii.  Both clusters were $10^5 \msun$ with 
$R_{\rm S} = 3$ pc and initially in 
virial equilibrium and the gas was lost gradually starting at 5 Myr over a 
timescale of 5 Myr.}\label{figure:diffrg}
\end{figure}

In fig.~\ref{figure:diffrg} the escape rate of particles from clusters 
at four different 
Galactocentric radii is illustrated.  The severity and time of onset 
of mass loss is significantly decreased with higher Galactocentric radii.  
The cluster at 2 kpc starts losing 
particles even before the onset of mass loss (at 5 Myr) as the cluster 
moves towards a dynamical equilibrium.

\begin{figure}
\centerline{\psfig{figure=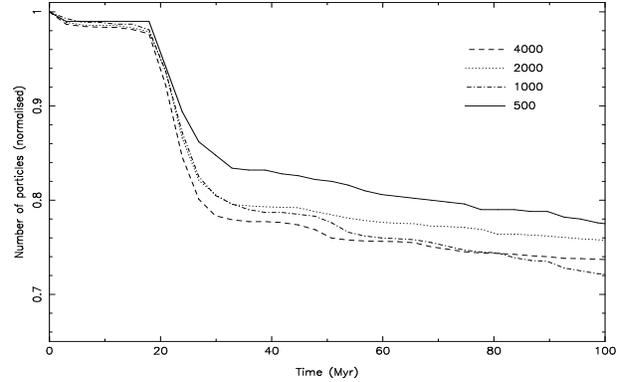,height=5.0cm,width=8.0cm,angle=270}}
\caption{The change of the number of particles (normalised to 1 at $T=0$ Myr) 
with time for runs with 
four different numbers of particles: 500 (full line), 1000 (dot-dashed line), 
2000 (dotted line) and 4000 (dashed line).  All four 
clusters were set up to represent the same physical cluster of $10^5 \msun$, 
$\alpha = 3.5$, $R_{\rm S} = 5.9$ pc, $R_{\rm G} = 5$ kpc with an SFE of 50\% 
and gradual early mass loss.}
\label{figure:particle}
\end{figure}

The scaling and properties of any $N$-body simulation vary with the 
number of particles used.  The selection of computational parameters is 
such as to try and reduce these effects and provide as good an 
approximation to reality as possible.  Figure~\ref{figure:particle} shows 
the change in the number of particles of a typical run as the number of 
particles used were varied from 500 to 4000.  This particular set of 
parameters were chosen as they result in a disrupted cluster when compared 
with the constraints of CS (see section 3.3).  As can be 
seen within this range the results of the change in particle numbers 
remain fairly consistent.  The results diverge more after about 200 Myr 
where a lower number of particles tends to lead to disruption far faster 
than a larger number.  However, it is only the first 50 Myr, or so, that 
these simulations cover, in which time the loss rates are comparable.

Figure~\ref{figure:particle} also provides an idea of the noise present 
in any one individual run in the mix of the curves and the lack of an 
obvious trend with particle number at these early times.  
To test the validity of the results generally, most runs were repeated 
at least once (usually several times) with slightly varying initial 
conditions.  A few of the border-line cases were run 10 or 20 times, this 
process has the advantage of testing the statistical robustness of the 
results.  The results from these varied runs were not found to differ 
significantly, hopefully showing that the changes observed for different 
sets of initial conditions were not the results of statistical 
fluctuations in the computations (see Heggie 1995).

\subsection{The effects of the three gas loss mechanisms}

The effect of the particular gas loss mechanism is found to be relatively 
consistent, independent of the mass or Galactocentric radius of the cluster, 
subject to the caveats mentioned in the previous section.  The three 
different mechanisms of gas expulsion are found to have quite different 
effects upon the stability of the cluster.

In this subsection all the clusters described are $10^{5} \msun$ at a 
Galactocentric radius of 10 kpc with an IMF slope of $\alpha = 3.5$ and 
Plummer model scale lengths initially $R_{\rm S} = 4.4$ pc (corresponding to a 
mean harmonic particle radius of 7.5 pc), with 
central stellar densities of $\approx 280 \msun$ pc$^{-3}$.      
This set of parameters is chosen as an example to illustrate the general 
effects of the gas expulsion mechanisms.  The clusters were all initially in 
virial equilibrium.

\begin{figure}
\centerline{\psfig{figure=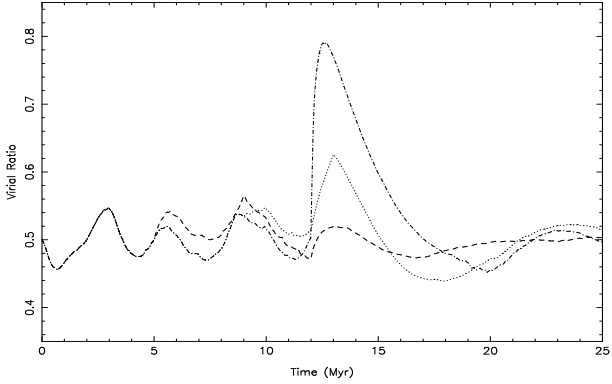,height=5.0cm,width=8.0cm,angle=270}}
\caption{The change of the virial ratio due to the three mechanisms 
of residual gas expulsion.  Early gradual loss beginning at 5 Myr and 
lasting for 5 Myr (dashed line).  Late gradual loss beginning at 9 Myr and 
also lasting for 5 Myr (dotted line). And a supershell starting at 12 
Myr (dash-dot line).  The three clusters were initially identical with 
$M_{\rm tot} = 10^5 \msun$, $\alpha = 3.5$, $R_{\rm S} = 4.4$ pc at a 
Galactocentric distance of 10 kpc.}
\label{figure:virial}
\end{figure}

Figure~\ref{figure:virial} shows the change in the 
virial ratio $Q=T/\mid \Omega \mid$ for three   
otherwise identical clusters each with a different mechanism of gas  
expulsion.

\begin{figure}
\centerline{\psfig{figure=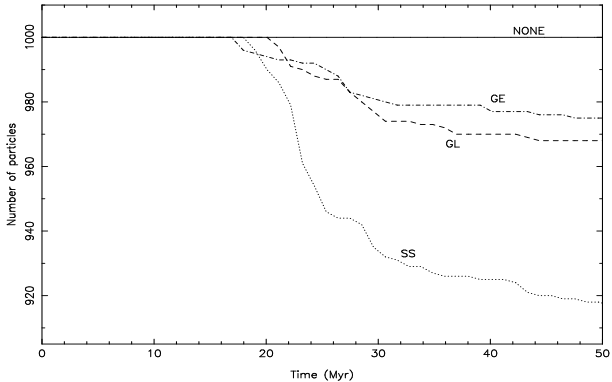,height=5.0cm,width=8.0cm,angle=270}}
\caption{The change in particle number of four initially identical $10^5 \msun$ 
clusters for each of the three gas loss mechanisms.  The 
lines represent: no gas loss (NONE) - solid line;  gradual early expulsion 
(GE) -  dash-dot line; gradual late expulsion (GL) - dashed line; and a 
supershell (SS) - dotted line.  The timescales for gas loss are the same as 
in fig before.}
\label{figure:diffmech}
\end{figure}

Figure~\ref{figure:diffmech} shows the loss of particles with time for 
each of the different residual gas expulsion mechanisms illustrated in 
fig.~\ref{figure:virial} as well as 
an identical cluster but with no gas present to be expelled.  
As can be seen from fig.~\ref{figure:diffmech} no loss of the stellar 
mass of the cluster occurs where there is no residual gas present to 
expel (in other sets of initial conditions some particle loss may occur 
but it is not significant in clusters that CS predict to survive).

Gradual early mass loss  is the least disruptive mechanism for the 
expulsion of gas.  
The changes in the potential caused by the gas over a period of 
$\approx 5$ Myr is small in comparison to the crossing time, and 
allows stars to adjust gradually.  In fig.~\ref{figure:virial} it 
can be seen that the system 
remains in approximate virial equilibrium for the duration of the 
gas loss.  Gradual late gas expulsion is found to be more disruptive.  This 
occurs as the delay (in this case until 9 Myr) causes the gas expulsion and 
mass loss from the supernovae of the most massive stars to be 
simultaneous (as, indeed, they should be as it is the supernovae that are 
assumed to be driving the gas expulsion).  This increases the mass 
loss from the cluster at any time above that of the gradual early 
mechanism resulting in a more disruptive effect upon the cluster.  Both of 
these mechanisms of gas loss show a 
'kink' in their virial ratios at $\approx 12$ Myr.  This is caused by the 
further disruptive effect of the loss of stellar mass when the most massive 
star particle begins to lose mass to simulate supernovae in the code.  As 
can be seen for the virial ratio of the early gradual loss simulation, 
this effect, in itself, is not particularly disruptive.  This is further 
supported as it does not induce particle loss in the cluster without 
any gas loss (fig.~\ref{figure:diffmech}).  

Gas loss via a supershell is the most disruptive mechanism 
of expulsion.  Most stars in a cluster are in the inner few pc of a cluster, 
typical half mass radii range from a few pc for low Galactocentric radii 
to several pc for larger Galactocentric radii.  A supershell will 
sweep the inner few pc of a cluster of its gas in only a few tens, possibly 
hundreds, of thousands of years.  This corresponding change in the 
potential in these central regions then occurs on a timescale less 
than a crossing time.  With such a short timescale, the stars in the cluster 
are not able to adjust their orbits to the new potential.  This 
sudden change pushes the cluster far out of virial equilibrium with 
$Q_{\rm max} \approx 0.8$ (fig.~\ref{figure:virial}).  As can be seen, 
however, the cluster is able to very rapidly readjust to the new potential 
and bring itself back into virial equilibrium very rapidly (within $\approx 
10$ crossing times).

It should be noted that while the supershell mechanism induces far more 
mass loss than either of the other two mechanisms it may not, necesserally, 
be too much more disruptive.  At higher Galactocentric radii and masses 
a larger and larger bound central mass remains which appears as if it 
could well survive.  A relatively small decrease in the scale length of the 
Plummer model can vastly reduce the level of mass loss from a supershell 
and prevent its disruption (see section 3.3).  

In fig.~\ref{figure:diffmech} the particle loss can be seen to be 
delayed by 10 to 15 Myr 
after the initiation of residual gas expulsion.  This delay is due to the 
travel time of particles with escape velocity from the cluster centre 
to the tidal radius at which point they are removed. 

During the gas loss episode the globular cluster expands from its 
initial central distribution into a new equilibrium distribution.  
In the settling process after star formation (see section 2.3) 
some stars will escape beyond the tidal 
radius in a slow process that will continue for the life of the cluster.  
This process is, however, enhanced and encouraged by a gas loss episode.  
The more rapid the gas loss, the more pronounced is the stellar mass 
loss associated with that gas loss.

\subsection{Survivability and gas loss from 50\% SFE globular clusters}

In this sub-section the effects of residual gas expulsion from clusters 
with an SFE of 50\% is investigated.  The state of the clusters 
after the gas expulsion episode is then compared to the survivability 
estimates of CS (fig.~\ref{figure:cs}).  

The effect of the residual gas expulsion described in the previous two 
sub-sections is to weaken the binding of a cluster.  In some cases 
this almost immediately (within a few hundred million years at most) results in 
the disruption of a cluster that, according to the estimates of CS, 
would have survived for a Hubble time without the gas expulsion.  In other 
cases the effect of the gas expulsion is far slower, but will still 
eventually result in the disruption of the cluster. 

\begin{figure}
\centerline{\psfig{figure=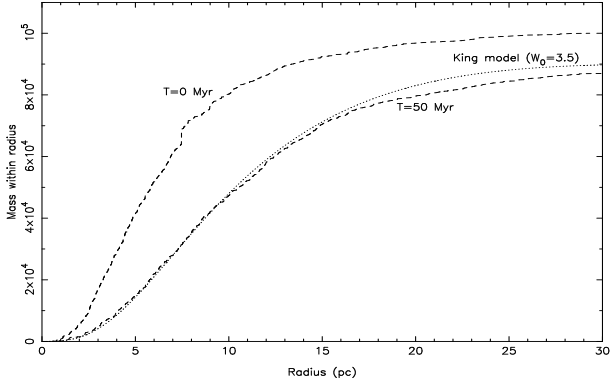,height=5.0cm,width=8.0cm,angle=270}}
\caption{The change in the mass distribution with radius 
during the time of gas expulsion.  The two dashed lines show the mass 
distribution of the simulation at 0 and 50 Myr.  The dotted line is 
the mass distribution expected of a King model with $W_{0} = 3.5$ (the 
lowest surviving King model with these initial conditions from 
CS).}\label{figure:king1}
\end{figure}

Figure~\ref{figure:king1} shows the effect of residual gas loss upon 
the distribution 
of mass with radius inside a cluster.  This cluster has a stellar mass of 
$10^5 \msun$ initially, an IMF slope of $\alpha = 3.5$ at a Galactocentric 
radius of 5 kpc.  The cluster has a initial Plummer length scale 
$R_{\rm S}=4.4$ pc, which gives a central stellar density initially of 
$\approx 280 \msun$ pc$^{-3}$.  The gas expulsion is modelled as the 
least disruptive, gradual early loss via stellar winds and the UV flux.  
After 50 Myr (which includes the residual gas expulsion and the evolution 
of the most massive stars) its mass distribution and rms velocity distribution 
are compared with those of the lowest King model that can CS estimate can 
survive for a Hubble time with the same initial mass, Galactocentric radius and 
IMF as the model cluster.  The use of a radial mass distribution 
is justified by the approximate sphericity of the clusters.  The initial 
conditions produce spherically symmetric clusters without rotation that 
retain their sphericity at least for the duration of the simulations.  
Variations in the sphericity are seen of a few percent but these fluctuations 
are not found to be in any way systematic.

\begin{figure*}
\centerline{\psfig{figure=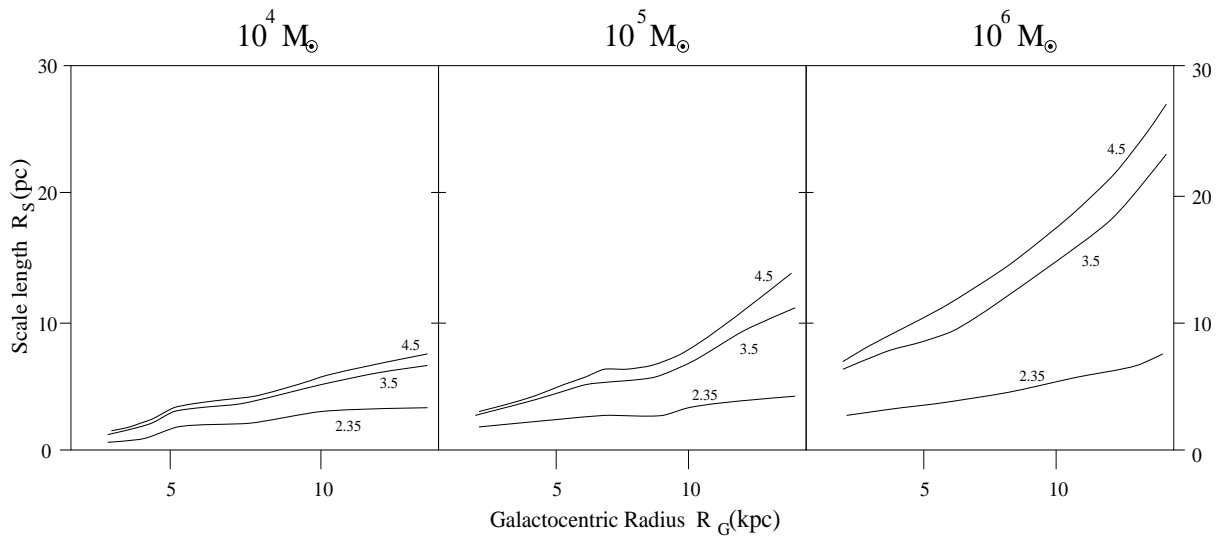,height=7.0cm,width=16.0cm,angle=270}}
\caption{The minimum length scale of a Plummer model required for the 
survival of a cluster of the indicated mass.  The IMF slope, $\alpha=$ 
2.35, 3.5 or 4.5 of the simulation are indicated on each line.  If a 
cluster is to survive it must lie below the line.}
\label{figure:results}
\end{figure*}

The King models used for comparison were numerically integrated from 
King (1966) and set to be similar to the initial condition King models used 
in the $N$-body calculations of Fukushige \& Heggie (1995).  They were given a 
$W_{0}$ equal to the lowest surviving value taken 
from fig.~\ref{figure:cs} and a total mass equal to the mass of the 
cluster at that time (which may significantly lower than the initial 
mass due to particle loss).  However, at lower masses 
fig.~\ref{figure:results} indicates that survival is actually easier, so this 
effect may underestimate survivability slightly.

\begin{figure}
\centerline{\psfig{figure=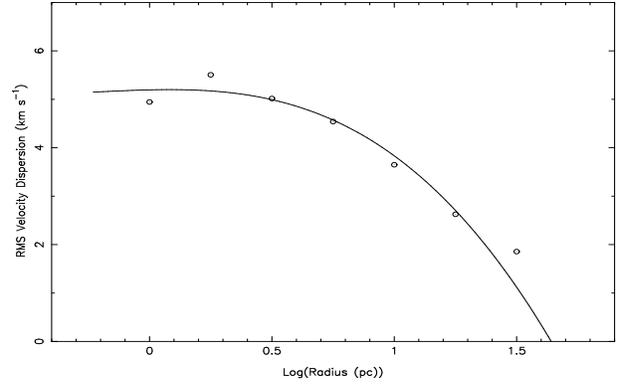,height=5.0cm,width=8.0cm,angle=270}}
\caption{The rms velocity dispersion of the same model cluster as 
in fig.~\ref{figure:king1} binned with 
radius at $T=50$ Myr (dotted circles) compared to the rms velocity 
dispersion of an $W_{0}=3.5$ King model of the same mass (solid  
line).}\label{figure:rms}
\end{figure}

Figure~\ref{figure:king1} shows that a cluster initially well within the 
survivability range of fig.~\ref{figure:cs} will, in this border-line 
case, fall to become a cluster only just expected to survive.  The cluster 
has lost nearly 10\% of its mass through both stellar evolution ($\approx 1\%$) 
and the escape of particles ($\approx 8\%$).  The 'jump' in 
the $T=0$ Myr mass profile at $\approx 7$ pc is caused by the 
large particle that represents the most massive stars before they have 
evolved.  As explained in section 2.2, the presence of a particle of this 
size for a short period does not significantly effect the dynamics of the 
simulation.  The similarity of the state of the cluster after 50 Myr 
to the $W_{0}=3.5$ King model is further illustrated in 
fig.~\ref{figure:rms}.  The rms velocity dispersions of the simulation 
can be seen to be very similar to those calculated for the fitted 
King model.  The dissimilarity at large radii is largely due to the 
low number of particles at these radii with which to construct the 
velocity dispersion.
The similarity of clusters to King models in these simulations is 
striking, even shortly after a disturbing episode of gas expulsion.

\begin{figure}
\centerline{\psfig{figure=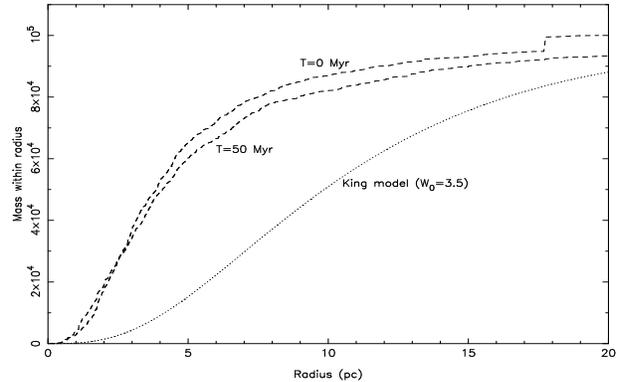,height=5.0cm,width=8.0cm,angle=270}}
\caption{The change in the mass distribution inside a certain radius 
for a cluster with no residual gas.  The two dashed lines show the mass 
distribution of the simulation at 0 and 50 Myr.  The dotted line is 
the mass distribution expected of a King model with $W_{0} = 3.5$ (the 
lowest surviving King model with these initial conditions from 
CS).}\label{figure:king2}
\end{figure}

It should be noted that the lowest survivability King models from CS 
are those before the most massive stars in the cluster evolve.  In these 
comparisons the simulations have already lost the most massive ($M > 
8 \msun$) stars.  The change in the mass distributions of clusters 
with no residual gas (and hence no gas expulsion) is shown 
in fig.~\ref{figure:king2} for a cluster with the 
same initial conditions as that in fig.~\ref{figure:king1}.  The range 
in radius has been reduced in this figure to show more clearly the 
increase in the central density of this simulation when compared 
to those with residual gas expulsion.  
The settling of the cluster and the evolution of the most massive 
stars can be seen to produce a slight lowering of the mass 
profile in the outer regions.  The King model illustrated is, again, the 
lowest surviving King model from fig.~\ref{figure:cs} this time 
with a mass of $9.6\times 10^{4}\msun$ as the cluster has lost 
$4\times 10^{3}\msun$ due to the stellar evolution of the most 
massive stars but no particles have escaped in this case.  The $T=0$ Myr 
'jump' in the mass profile at $\approx 18$ pc is, again, caused by the 
presence of the massive star particle.

This process of estimating the survival, or otherwise, of a cluster 
was followed for a wide range of initial conditions similar to those 
in CS.  Comparisons were made after the 50 Myr evolutionary period to 
try and determine if the cluster would still be bound after a Hubble 
time.   

It is found, unsurprisingly, that the minimum concentration required for 
the survival of a cluster is always increased.  This increasing concentration 
in terms of a Plummer potential corresponds to an increasing $W_{0}$ of 
King models for that initial mass and Galactocentric distance.

The preferential survival of clusters at lower Galactocentric radii is more 
than compensated for by the increased disruptive effects of residual 
gas expulsion at those radii.  The lowering of the required King model 
at low Galactocentric radii becomes an upturn, leading to a higher 
concentration for survival at low Galactocentric radii.

Figure~\ref{figure:results} shows the effects of residual gas expulsion on 
the survivability of globular clusters.  Figure~\ref{figure:results} is 
laid-out in a similar fashion to fig.~\ref{figure:cs} for the three 
different initial stellar masses and IMF slopes.  It should be noted that 
in fig.~\ref{figure:results} the steepest slope to the IMF is the uppermost 
line and clusters that survive are below the lines (the opposite of 
fig.~\ref{figure:cs}) this is due to the choice of scale length as the 
abscissa.  It should be noted that all 
of these clusters are in virial equilibrium, the effects of non-equilibrium 
initial conditions are discussed below.  

The results are given in terms of the length scale, $R_{\rm S}$, 
of the Plummer model for the stars used as the initial conditions.  The 
length scale can be converted into a central density via the formula 

\begin{equation}
\rho_{0} = \frac {3M_{\rm tot}}{4\pi R_{\rm S}^{3}}
\end{equation}

\noindent  An interesting feature of a conversion to central densities 
is that the central densities required for survival are dependent only 
upon the IMF slope and the Galactocentric radius, and are approximately  
independent of the initial mass of the cluster.  An order of magnitude 
increase in the initial mass causes an increase in the scale length 
required for survival.  When $\alpha = 3.5$ and 4.5 this increase is 
approximately a doubling of the required $R_{\rm S}$.  For $\alpha = 2.35$, 
however,  the relationship is not as clear and depends more upon 
Galactocentric radius.  The increase required ranges from $\approx 30\%$ 
at low Galactocentric radii to $100\%$ at higher radii.  This relationship 
is not present in the results of CS and must therefore appear as a property 
of the residual gas expulsion.  Although this result depends upon the slope 
of the IMF, it appears to hold for each of the three different values 
of $\alpha$.

The clusters represented in fig.~\ref{figure:results} have all undergone 
the gradual early gas expulsion, the least disruptive of the three expulsion 
mechanisms.   The quoted values for $R_{\rm S}$ in fig.~\ref{figure:results} 
should therefore only be thought of as upper limits for the gas expulsion 
mechanisms. 

The change in the length scale required for survival produced by each of 
the other two gas expulsion mechanisms is not very sensitive to IMF slope 
or Galactocentric radius.  Gradual late gas expulsion 
reduces the required $R_{\rm S}$ from gradual early expulsion by around 
4 or 5\%, while a supershell expulsion will reduce $R_{\rm S}$ by 
about 8 to 10\%.  This has the effect of lowering the position of the lines 
on fig.~\ref{figure:results} and hence survival requires a slightly 
higher central density.  As the reductions 
are similar for all Galactocentric radii and IMF slopes, the independence 
of the required central density to mass is retained for all gas expulsion 
mechanisms.  

Even if gradual early mass loss was not able to completely remove the 
residual gas from a cluster, its action may reduce the disruptive effects of 
one of the latter mechanisms.  If the stellar winds and UV flux were able 
to remove some of the gas, or even just decrease the density of the 
residual gas in the central few pc of the cluster, then the latter 
mechanisms would not be as disruptive. 

The cut-off line on fig.~\ref{figure:results} should be considered merely as a 
guide to the actual position of the cut-off between survival and disruption.  
Many of the simplifying assumptions made in this paper and in CS mean that 
it is only a first order estimate of the true value which will depend on 
many initial properties individual to any particular cluster.

The quoted value of the mass is the total initial stellar mass of the 
cluster.  After gas loss the cluster will, of course have lost 50\% of its 
initial mass after losing the residual gas as all of these clusters have 
an SFE of 50\%.  In addition, some clusters that survive can loose from 
5\% to 20\% of their initial stellar mass by stellar evolution and the 
escape of particles.  This loss should not effect the survivability of 
the cluster after residual gas expulsion as lower mass clusters are 
more stable after residual gas expulsion than higher mass clusters (as 
illustrated in fig.~\ref{figure:cs}). 

All clusters on the border-line of survivability lose mass during the 
gas expulsion episode and the restabalisation period afterwards.  The 
extent of this mass loss appears to be relatively independent of the 
Galactocentric radius of the cluster.  The increase in $r_{\rm t}$ acts to 
cancel the increase in the border-line value of $R_{\rm S}$. The mass 
loss is still highly dependent upon the initial mass of the 
cluster, however.  $10^6 \msun$ clusters only lose a 
few percent of their mass, the vast majority of this mass loss being the 
mass lost in the stellar evolution of the most massive stars.  $10^4 
\msun$ clusters, however, may lose up to 25\% (typically 10 to 15\%) of 
their mass during this phase.  The reason for this appears to be the 
greater extent of the gravitational influence of the larger clusters.  The 
tidal radius of a $10^6 \msun$ cluster will be approximately 5 times larger 
than that of a $10^4 \msun$ cluster at the same Galactocentric radius.  
The more massive clusters are then more capable of retaining a weakly 
bound extended halo of stars than those of lower mass.  These extended 
halos will presumably be stripped relatively quickly by the Galactic 
tidal field, but in these simulations as they are still within the 
tidal radius, they are still counted as being part of the cluster.  Due 
to this effect (and the lack of particles at large distances), the 
comparisons to King models were concentrated on the inner regions (usually 
within the half-mass radius) of clusters where they are expected to be 
more accurate indicators.
In the vast majority of cases, 50\% SFE clusters were found to be very 
similar to King models after gas expulsion.

\subsection{Non-virial equilibrium clusters.} 

The assumption that the stars within a cluster are in virial equilibrium 
at formation has underlay the previous discussion of the survivability 
of a globular cluster.  The validity of this assumption, however, is 
not known.  The gas from which the stars form is expected to be in virial 
equilibrium itself in globular cluster formation models with internally 
induced star formation.  These stars may then be expected to also be in, 
or very close to virial equilibrium.  If star formation is induced by 
external effects then there seems no a priori reason to expect the 
stars themselves to be in equilibrium.  On the contrary, it may be 
expected that the stars would not be virialised.

\begin{figure}
\centerline{\psfig{figure=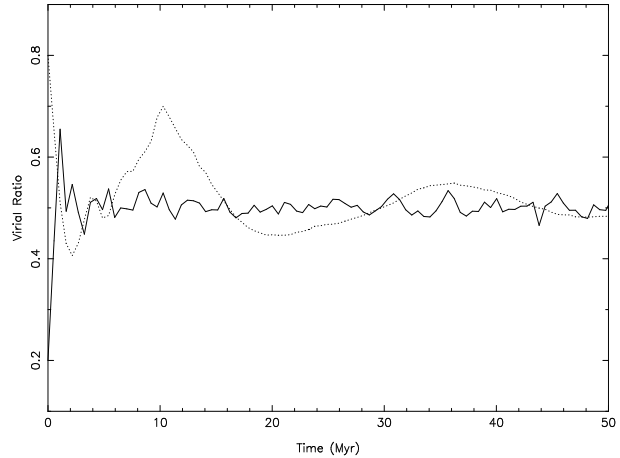,height=6.0cm,width=8.0cm,angle=270}}
\caption{The evolution of the virial ratio $Q$ for two $10^5 \msun$ clusters 
with $R_{\rm S} = 4.4$ pc, $\alpha = 3.5$ at 5 kpc.  The evolution shown by 
the dotted line is that of a cluster with an initial virial ratio of $Q=0.8$, 
while the solid line is for a cluster with initial virial ratio $Q=0.2$. 
The clusters both underwent the same form of gradual early mass loss 
beginning at 5 Myr, and lasting for 5 Myr.}
\label{figure:diffvir1}
\end{figure}

Figure~\ref{figure:diffvir1} shows the evolution over 50 Myr of two 
clusters with initial virial ratios of $Q=0.2$ and $Q=0.8$.  The $Q=0.2$ 
cluster can be seen to settle into virial equilibrium very quickly (within 
5 Myr), the subsequent variations in the virial ratio being the expected 
random level of fluctuation.  The cluster initially at $Q=0.8$, however, is far 
more disturbed.  The action of attempting to settling into a virial 
equilibrium distribution combined with residual gas expulsion and mass 
loss due to stellar evolution in the first 10 Myr of the cluster's 
evolution prevent the cluster from reaching equilibrium in the 50 Myr 
shown.  The cluster is in the process of attaining virial equilibrium, but 
the loss of energetic stars required to cause this has been extreme.

\begin{figure}
\centerline{\psfig{figure=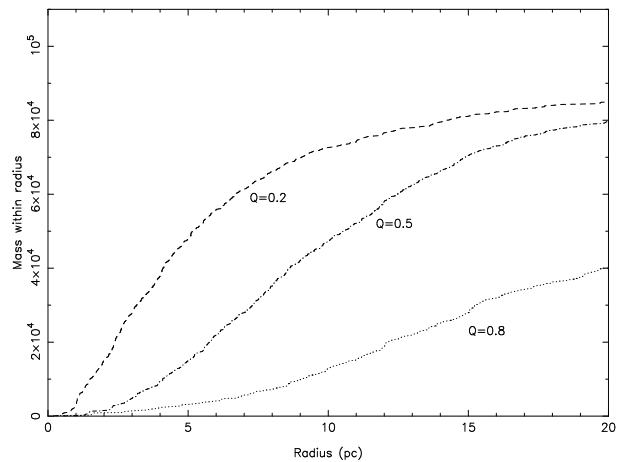,height=6.0cm,width=8.0cm,angle=270}}
\caption{The mass distribution of three otherwise identical clusters with 
different initial virial ratios, $Q=0.2, 0.5$ and $0.8$, after 50 Myr with 
gradual early mass loss.  The Q=0.5, $R_{\rm S} = 4.4$ pc cluster is at 
the limit of survivability for a $10^5 \msun$ cluster with $\alpha = 3.5$ at 
5 kpc.}
\label{figure:diffvir2}
\end{figure}

In a cluster where gas expulsion (by the gradual early mechanism) is 
included the $Q=0.2$ cluster has lost exactly 10\% of its 
stellar mass after 50 Myr 
(exactly the same mass loss as from an equivalent cluster starting 
with $Q=0.5$).  The $Q=0.8$ cluster has, however, lost nearly 40\% of its 
initial stellar mass in the same period.  Figure~\ref{figure:diffvir2} 
shows the effects that the different initial conditions have had upon the 
final mass distributions of the cluster.  A cluster with the same 
initial conditions, but initially in virial equilibrium is a border-line 
case for surviving a Hubble time.  As is illustrated, a different 
initial virial ratio will, in the case of $Q<0.5$, improve the survivability 
of the cluster or, in the case of $Q>0.5$, substantially decrease the 
chances of surviving for a Hubble time.  The main cause of this effect 
appears to be the changes in concentration that occur within a cluster 
out of virial equilibrium in order to achieve virial equilibrium:  $Q<0.5$ 
clusters will collapse and $Q>0.5$ clusters expand.  These changes 
effectively increase or decrease the length scale, shifting the cluster 
down or up fig.~\ref{figure:results}. 

\subsection{Survivability and gas loss from less than 50\% SFE globular clusters}

The clusters discussed in the previous sections have all had an SFE of 
50\%.  Considering the low SFEs observed in star forming regions 
in the Galaxy today, only of order a few percent (Larson 1986), it is of 
interest to examine the effects 
of the loss of residual gas from clusters that have only achieved an 
SFE of less than 50\% and to estimate the survivability of those clusters.  

Lada \et (1984) followed a similar method in an $N$-body simulation of the 
effect upon star clusters of residual gas expulsion.  They found that 
star clusters may remain bound with an SFE of 30\%, however they will 
lose 10 to 80\% of their original stars in this process.  These 
simulations differ from the simulations presented within this paper in 
the size of the cluster that is simulated, the timescale of the gas 
expulsion and the number of particles used. 

The effect of residual gas expulsion from a cluster with an SFE lower than 
50\% is, unsurprisingly, qualitatively no different from that from the 
clusters described above with an SFE of 50\%.  Quantitatively, an SFE of 
less than 50\% decreases the scale length of the lowest surviving Plummer 
model as its effect upon the cluster is more disruptive.  There 
would appear, theoretically, to be no lower limit to the SFE that will 
produce a bound cluster, the required central density would just continue to 
increase with decreasing SFE as long as the tidal radius is large 
enough to contain the expansion of the cluster after the residual gas 
expulsion.  In practice, however, very low SFEs require a 
concentration and central density that would be highly implausible, if not 
impossible to achieve.  

The effective limit of cluster SFEs occurs when a 
cluster does not contain enough massive stars to expel the residual 
gas.  Following Morgan \& Lake (1989) and using equation (7) it is possible 
to obtain a rough estimate of the minimum SFE for a particular mass 
and IMF slope.  The following table shows the minimum values of the SFE 
allowed if the cluster is to be able to expel its residual gas via 
supernovae alone.  

When $\alpha = 2.35$ clusters of all masses contain 
enough supernovae to expel residual gas down to any plausible value of 
the SFE (of fractions of a percent).  For $\alpha = 4.5$, however, the 
required SFE rises to very high values as the number of massive stars 
in a cluster falls.  In a $10^4 \msun$ cluster, for example, equation (7) 
predicts only 0.05 stars $\gte 8 \msun$, hence the '?' in that 
position.  The strong dependence of the minimum SFE upon mass occurs as 
the number of massive stars is proportional to the mass, while the 
number of supernovae required to disrupt a cloud is proportional to 
$M_{\rm cl}^{5/3}$.

These values should be taken as upper limits only as the calculations 
of Morgan \& Lake do not include energy input to the residual gas 
prior to the supernovae and they do not account for complex hydrodynamic 
effects within the gas.  Even so, they are suggestive that IMF slopes 
of $\alpha \gte 4.0$ are implausible for low or moderate SFEs.  It also 
suggests that gradual early mass loss would probably  be ineffective 
at expelling the residual gas, or even reducing its density in the 
central regions of the cluster for low SFEs.  In that case the more disruptive 
gradual late and supershell expulsion mechanisms would be required with their 
additional constraints upon survivability.

For SFEs as low as 30\% comparisons with CS to assess survivability are 
still thought to be fairly valid as after the gas expulsion episode the 
clusters retain a good resemblance to King models, especially within the 
half mass radius.  Out side of the half mass radius the mass density and 
rms velocity dispersion often fall far more rapidly than a King model.  
For this reason survivability becomes far more difficult to asses 
by this method.  There does appear to be a linear decline of the 
scale length (hence an increase in the central density) required for 
survival.  For a 50\% SFE $10^5 \msun$ cluster at $R_{\rm G} = 5$ kpc 
the border-line occurs at $R_{\rm S} \approx 4.4$ pc.  With an SFE of 
40\% this falls to $R_{\rm S} \approx 3.3$ pc and for an SFE of 30\% to 
$R_{\rm S} \approx 2.9$ pc.

In clusters with less than approximately a 30\% SFE the resemblance of the 
clusters to King models after the gas expulsion episode is not good.  For 
that reason, estimates of the survivability by the method of comparison to 
CS are not expected to be accurate in these cases.  Proper estimates for the 
survivability of these clusters (by running the code for time periods 
comparable to a Hubble time) are beyond the scope of this present paper.  Such 
a simulation would have to address a number of the problems associated 
with $N$-body simulations that this paper has attempted to minimise in the 
treatment of the clusters over a short time (with respect to any significant 
dynamical timescale).

\begin{center}
\begin{table}
\caption{The lowest SFE required for a cluster to contain enough massive 
stars ($\protect\gte 8 \msun$) to expel its residual gas for each of the 
three slopes to the IMF.  From the calculations of Morgan \& Lake (1989).}
\begin{center}
\begin{tabular}{|c|c|c|c|} \hline
 & $\alpha = 2.35$ & $\alpha = 3.5$ & $\alpha = 4.5$  \\ \hline
$10^4 \msun$ & 1\% & 11\% & ?  \\ 
$10^5 \msun$ & $>$ 0\% & 20\% & 75\% \\ 
$10^6 \msun$ & $>$ 0\% & 56\% & 90\% \\ \hline
\end{tabular}
\end{center}
\end{table}
\end{center}

The clusters tended to show a more concentrated core, well 
within the half-mass radius with a far larger number of stars escaping 
and forming a loosely bound or unbound halo around the cluster before 
being stripped by the Galactic tidal field.  The existence of this concentrated 
core may indicate the possibility of forming a bound cluster of 
significantly lower mass than the original cluster (cf. Lada \et 1984).  
The linear decline of the required scale length with SFE appears to 
continue below an SFE of 30\%, due to the uncertainty in the validity 
of the application of the comparisons at these SFEs this result has very 
little weight.  What does appear clear, however, is the very real 
possibility that these clusters could survive (albeit largely depleted) 
with such a low SFE.

\section{Discussion}

This paper explored the effects of an episode of residual gas expulsion 
a few Myr after star formation in a globular cluster.  This episode 
clears the cluster of all the gas not turned in to stars.  An $N$-body 
code, based upon Aarseth's nbody2 code, with a variable external 
potential representing gas acting upon 
the stellar particles was used to perform a large number of simulations 
over a wide variety of initial conditions (IMF slope, Plummer scale 
length, SFE, virial ratio and Galactocentric radius).  The results 
of these simulations after 50 Myr were compared to the results of Chernoff \& 
Shapiro (1987) to provide an estimate of the survivability of the 
globular cluster.  The short timescale of the $N$-body simulations 
helps to reduce the errors inherent in $N$-body calculations.

The residual gas loss from globular clusters was modelled using three 
idealised cases.  All three mechanisms of gas expulsion rely upon the 
input of energy in to the cluster gas by massive stars to provide the 
impetus for the expulsion.  The first mechanism assumes that the UV 
flux and stellar winds from the most massive stars will gradually force the 
residual gas out of the cluster and beyond the tidal radius where it 
will be lost into the general Galactic environment.  The second mechanism 
follows the same path of gradual expulsion but is caused by the supernovae 
of these massive stars.  Both of these mechanisms involve the reduction 
of the mass of gas in the external potential over a timescale of a few 
Myr.  The third mechanism, however, is based upon the expulsion of the 
gas in a 'supershell' formed by the merging of the shock fronts of 
many supernovae in the central regions of the cluster.  This mechanism 
sweeps the cluster clean of gas on a timescale dependent upon the number 
of supernovae and the mass and distribution of the gas.

The imposition upon a globular cluster of the condition that it must expel 
its residual gas soon after star formation can be seen to place strong 
constraints upon the set of initial conditions that will result in a 
cluster being able to survive for a Hubble time.  As illustrated in 
fig.~\ref{figure:results} these constraints relax in an approximately 
linear way with increasing Galactocentric radius.  

There are two schools of thought as to the extent of the original 
globular cluster population.  The first suggests that the present population 
of globular clusters is a fairly complete sample of the original population 
(some of these arguments are summarised in Laird \et 1988).  If this is 
the case then all proto-globular clusters must have formed in such a 
way as to survive residual gas loss, dynamical evolution and destructive 
processes.  In such a case the initial conditions implied by these 
simulations would represent a lower limit upon those of actual globular 
clusters.  Aguilar, Hut \& Ostriker, however, conclude from their 
study of the destructive mechanisms that operate upon globular clusters 
that the current population may only be '...but a shadow of its former 
self.'  Even in the case where a large number of globular clusters are 
destroyed by the Galaxy, there is a lower limit of $\approx$ 2\% of the 
initial population surviving, as globular clusters still comprise 
2\% of the visible halo mass.  Whatever the case may be, then, a 
significant number of original globular clusters are expected to 
survive their residual gas expulsion phase.

At the low Galactocentric radii ($R_{\rm G} < 6$ kpc) that most globular 
clusters inhabit with an average mass of $10^5 \msun$ a central gas 
density in clusters just before the time of star formation of around  
$10^{3} \msun$ pc$^{-3}$ is 
required (for an SFE of 50\%) corresponding to mean densities within these 
clouds of $\approx 1 \msun$ pc$^{-3}$.  As stated in section 2.6, 
the limits obtained in this paper should be regarded as lower limits 
only upon the initial conditions necessary for a cluster to survive.  
These densities are of the order of those observed in giant molecular clouds 
in the Galaxy today, although their cores, where star formation would 
be expected to occur, are far less massive than a globular cluster (Harris 
\& Pudritz 1994 and references therein). 

An interesting result to arise in this treatment of residual 
gas expulsion is that the central density requirement for survival 
is approximately independent of the initial mass.  It depends only upon the 
IMF, Galactocentric radius and residual gas expulsion mechanism 
employed.  If the formation mechanism of all globular clusters produces 
similar IMFs and the gas expulsion mechanism is the same then globular 
clusters of all initial stellar masses will be able to survive at 
all Galactocentric radii if the central density in the proto-cluster cloud 
is above some minimum value ($\approx 10^3 \msun$ pc$^{-3}$ for a 50\% 
SFE cluster).  As noted above this is not an unreasonable figure to 
place upon the central densities of a proto-cluster cloud.

Unfortunately, any original independence of central density and mass 
will have been eliminated in the present day Galactic globular cluster 
system by dynamical evolution which can drive up central densities 
by gravothermal instabilities (Spitzer 1987) or possibly reduce 
it in clusters that are tending towards disruption.  The best testing 
ground for this independence would be young globular clusters such 
as those around the LMC which have not had the time to significantly 
evolve dynamically.  
A detailed comparison of observations of young clusters to theoretical 
considerations of cluster survival is in preparation.

The higher values of the slope, $\alpha$, of the IMF seem to lie fairly 
close together in  fig.~\ref{figure:results}, while $\alpha=2.35$ 
provides far stronger constraints on the length scale.  This effect 
is due mainly to the results of CS were the substantial dependence upon 
$\alpha$ is due to the mass loss from stellar evolution at later times 
than these simulations follow, rather than the gas loss mechanisms which 
are relatively insensitive to the IMF of the stars.  Nevertheless, a 
cluster of $\alpha = 2.35$ requires a central density at least an order of 
magnitude higher than $\alpha = 3.5$ or 4.5 in order to survive.  However, 
the calculations presented in table 1 appear to indicate that very high 
values of the IMF slope are not allowed as they do not provide 
enough high mass stars ($M \gte 8 \msun$) to expel gas in any cluster 
with even a $\gte 50\%$ SFE.

Observed luminosity functions in clusters today appear to show that 
globular clusters have a low slope to their mass functions for the 
low masses of stars present in clusters today, observed mass function slopes 
appear as low as $\alpha \approx 2$ (Fahlman 1993).  The theoretical 
results presented here and elsewhere, however, show that a high slope 
to the IMF is advantageous for the 
survival of globular clusters.  High slopes to the IMF will reduce the mass 
loss due to stellar evolution that occurs during the lifetime of the 
cluster (Chernoff \& Weinberg 1990).       
A possible explanation of this apparent discrepancy in the 
observed and theoretically allowed values of the IMF slope is the action 
of dynamical evolution including mass segregation and the preferential loss 
of low mass stars which would act to flatten the IMF (Piotto 1993).  Such a change in both the observed and actual mass function has been 
observed in the Fokker-Planck calculations of Chernoff \& Weinberg (1990).   
It is not clear that this effect could cause the level of change 
required (reducing $\alpha$ by 1 or 2) within a Hubble time, although 
Piotto (1993) concludes that the IMF of globular cluster stars will 
be 'strongly modified' by dynamical evolution.  New observations of 
deep luminosity functions should help to clarify if there is indeed a 
discrepancy.  It may also be instructive to 
run both Fokker-Planck and $N$-body simulations with different forms of the 
IMF to assess the effects that this might have upon the dynamical evolution 
of a system.

The gas expulsion mechanisms simulated in this paper are meant to 
provide a first approximation to the most plausible routes by which the 
residual gas may be expelled.  They are, however, highly idealised and, as 
such, may not fully simulate the interaction of stars and gas during 
this episode, possibly leading to an over estimate of the disruptive 
effects of the gas expulsion.  More realistic hydrodynamic simulations 
would need to be run to fully understand the effects of stellar winds, 
UV flux and supernovae on such a dense medium as that presumably 
found within very young globular clusters.  

For low SFEs large numbers of massive stars are required to expel the 
residual gas which implies a shallow slope to the IMF.  Apart from the 
problems this poses in later evolution, where 
low $\alpha$ mitigate against survival, low SFEs may require the action 
of the more disruptive gas expulsion mechanisms to clear the cluster of 
residual gas.  In these cases the central densities required of a cluster 
in order to survive will be extremely high ($\gte 10^4 \msun$ pc$^{-3}$).  
The obvious compromise would be to form clusters with moderate 
SFEs ($\approx 50\%$) and mid-range IMF slopes ($\alpha \approx 3$).

The suggestion has been made that globular clusters have any gas within 
them routinely removed by the ram pressure as they pass through the disc 
(Faulkner \& Smith 1991).  This may well be an adequate method of removing the 
relatively small amounts of gas that may collect within a cluster when 
it is lost by stars in the process of stellar evolution.  However, it 
is not clear if this method could remove the large amount of residual gas 
presumed present and, if it could, that it would be any the less disruptive 
than the mechanisms described within this paper.  It would seem that 
the removal of any large amount of residual gas by whatever mechanism will 
have a significant disruptive effect upon a globular cluster.

It has been shown in section 3.4 that clusters that have an initial 
virial ratio below that of virial equilibrium are considerably more 
stable than those clusters initially in or above virial equilibrium.  The 
higher survivability of $Q<0.5$ globular clusters may provide some 
evidence in favour of externally induced star formation models of 
globular cluster formation (such as those of Lin \& Murray 1991, Murray \& Lin 
1992, Shapiro, Clocchiatti \& Kang 1992, Kumai, Basu \& Fujimoto 1993, 
and Lee, Schramm \& Mathews 1995) 
where stars may well form out of virial equilibrium.  The star formation 
in such models will probably not occur in such simple distributions (and 
almost certainly not spherically) and so a proper investigation of such 
models is beyond the scope of this present paper.  Although a limit may 
be placed on the types of star formation that is permitted by the 
consideration that they must appear as King models within only a 
few tens of Myr (the observational constraints from Elson, Fall \& Freeman 
1987).

The possible presence of dark matter with globular clusters could provide 
a stabalising influence.  It has recently been suggested by Heggie \& Hut 
(1995) that up to half of the mass present today in globular clusters 
may be unobserved.  If a significant amount of dark matter were present 
at the time of the burst of star formation in a globular cluster this 
would provide additional mass to help reduce the disruptive effects of 
a gas expulsion episode.  However, if , as Heggie \& Hut suggest, this 
unseen mass is composed mainly of stellar remnants, the presence of dark 
matter may be evidence of a large number of high and intermediate mass 
stars in the cluster at formation.  Such a presence of stars would only 
exacerbate the problems already posed by stellar mass loss during 
evolution by implying a low value of the IMF slope.

Of course, all of these problems would not be important if the star 
formation mechanism were able to produce star formation efficiencies 
greatly in excess of 50\%, perhaps approaching 100\%.  Our 
knowledge of both star and globular cluster formation is sufficiently 
limited that such a possibility can not be ruled out.  Even if this 
were possible (and it is not clear how any system could achieve such high 
SFEs) the limitations from simulations not including residual gas expulsion 
would still hold. 

The results presented in this paper can be briefly summarised as follows:

\begin{enumerate}

\item The loss of a significant fraction of a clusters mass during a 
residual gas expulsion will effect the structure of the cluster 
significantly, disrupting many clusters that would otherwise be expected 
to survive.  However, a globular cluster may still survive with SFEs 
as low as 20\% if the initial concentration of the cluster is high enough.

\item The survivability appears only to depend upon the central density 
of a cluster at formation, not upon its mass (for a given IMF slope, 
Galactocentric radius and SFE).

\item For moderate SFEs ($\approx 50\%$) the central density required for 
survival at low Galactocentric radii is around $10^3 \msun$ pc$^{-3}$, 
similar to the central densities seen in giant molecular clouds in the 
Galaxy today.

\item The survival of a globular cluster would appear to be a play-off 
between the SFE and IMF slope, $\alpha$.  The clusters that 
survive with attainable central densities (at least they are observed in the 
Galaxy today) with a moderate $\alpha \approx 3$ require SFEs $\gte 40\%$.  
Such a value for the SFE is still well in excess of those values 
observed in star forming regions today, but is far more reasonable than 
previously assumed values approaching 100\%.

\end{enumerate} 

\section{Aknowledgements}

I would like to thank my supervisor, R.\,J.\,Tayler for his advice and 
S.\,J.\,Aarseth for his help and allowing me to use his code.  Also 
many thanks to A.\,A.\,El-Zant for his help and B.\,J.\,T.\,Jones for his 
comments. S.\,P.\,Goodwin is a DPhil student at the University of Sussex 
currently in receipt of a PPARC grant.

\end{document}